\documentclass{statsoc}

\usepackage{color}
\usepackage[a4paper]{geometry}
\usepackage{amsmath}
\usepackage{mathrsfs}
\usepackage{graphicx}

\newcommand{\bs}{\boldsymbol}
\newcommand{\mb}{\mathbf}

\title[SVARO]{Bayesian Analysis of fMRI data with \\ Spatially-Varying Autoregressive Orders}
\author[Teng, M. {\it et al.}]{Ming Teng}
\address{Department of Biostatistics, University of Michigan,
Ann Arbor, Mi., USA.}
\author[]{Farouk S. Nathoo}
\address{Department of Mathematics and Statistics, University of Victoria,\\
Victoria, British Columbia,
Canada.}
\author[]{Timothy D. Johnson}
\address{Department of Biostatistics, University of Michigan,
Ann Arbor, Mi., USA.}
\email{tdjtdj@umich.edu}

\begin{document}

\begin{abstract}

Statistical modeling of fMRI data is challenging as the data are both spatially and temporally correlated. Spatially, measurements are taken at thousands of contiguous regions, called voxels, and temporally measurements are taken at hundreds of time points at each voxel. Recent advances in Bayesian hierarchical modeling have addressed the challenges of spatiotemproal structure in fMRI data with models incorporating both spatial and temporal priors for signal and noise. While there has been extensive research on modeling the fMRI signal (i.~e., the covolution of the experimental design with the functional choice for the hemodynamic response function) and its spatial variability, less attention has been paid to realistic modeling of the temporal dependence that typically exists within the fMRI noise, where a low order autoregressive process is typically adopted. Furthermore, the AR order is held constant across voxels  (e.g. AR(1) at each voxel). Motivated by an event-related fMRI experiment, we propose a novel hierarchical Bayesian model with automatic selection of the autoregressive orders of the noise process that vary spatially over the brain. With simulation studies we show that our model has improved accuracy and apply it to our motivating example. 

\end{abstract}

%\newpage

\section{Introduction}\label{intro}

In the analysis of functional magnetic resonance imaging (fMRI) data a key challenge is dealing with spatial and temporal correlation. The temporal correlation can arise from many sources, including scanner drift at very low frequencies, slow vascular/metabolic oscillations that are typically of moderate to low frequency, and some other sources of noise such as breathing and heartbeat. Simply ignoring these sources of autocorrelation may lead to increased false positive discoveries \citep{makni2006joint}. To deal with these issues, a variety of approaches have been proposed. One commonly used approach, namely ``prewhitening", works by estimating the temporal autocorrelation and then de-correlating the noise using the estimates \citep{bullmore1996statistical, locascio1997time}. Besides these stationary time series models, non-stationary $1/f$ models have also been proposed \citep{zarahn1997empirical,bullmore2004wavelets}. According to \cite{friston2000smooth}, prewhitening can produce an extraneous source of bias. Alternatively, a band-pass filtering technique known as ``pre-coloring'' can be applied to the data first, followed by statistical modeling that deals with the autocorrelation in the colored data. For a review and discussion of these approaches the reader is referred to \cite{woolrich2001temporal}. While high-pass filtering has proven to be beneficial in increasing the power of the statistical analysis, the low-pass filtering involved in coloring is considered controversial in that it tends to add autocorrelation into the data \citep{skudlarski1999roc, della2002empirical}.

While accurate temporal modeling is important for estimation of the fMRI signal and its standard error, traditional approaches apply a temporal model at each voxel, independently.  That is, they ignore spatial correlation. More specifically, this mass univariate approach, considered to be the classical approach to the analysis of fMRI data, includes a smoothing step involving a spatial Gaussian filter that is applied to the data first \citep{friston1995analysis}, followed by model estimation at each voxel, and then statistical inference is based on random field theory \citep{worsley1995analysis} which is applied to adjust for multiplicity in the spatial domain. While this approach remains the most common approach for analyzing fMRI data it has been criticized on a number of grounds. For example, the Gaussian kernel that is used to smooth the data has to be pre-specified and introduces artificial correlation into the data. In addition, this approach does not directly account for spatial correlation in the model. 

Partly as a result of these criticisms, Bayesian models with spatial structured priors have been proposed that allow for the calculation of posterior probability maps (PPM) for activation. This Bayesian approach to inference is based on an explicit spatial model and does not require smoothing the data with a Gaussian kernel nor does it require the use of random-field theory-based adjustments for multiplicity. A variety of spatial-temporal Bayesian models have been proposed. One model that is widely used and implemented within the SPM software is the GLM-AR (general linear model, auto-regressive) model \citep{penny2003variational,penny2005bayesian,penny2007bayesian}, The GLM-AR model assumes that the data can be decomposed into two sources of variability. The first source is the product of the design matrix for the fMRI experiment convolved with a hemodynamic response function (HRF) and experimental factors, and the second source represents temporally correlated noise that is modeled using a low-order AR structure. In addition, the regression coefficients and the autoregressive coefficients vary across space and are assigned spatial smoothing priors.  \cite{gossl2001bayesian} proposed a model where the data are decomposed into three sources, a spatial stimulus, a deterministic trend, and a white noise process. However, this modeling approach may not account for some higher frequency stochastic noise components. \cite{woolrich2004fully} assumed that the temporal noise arises from both large scale and small scale variation, and built a space-time simultaneously auto-regressive model that accounts for both scales of variation. Methods focusing on spatial variable selection have also been proposed (see, e.g., \cite{bezener2016bayesian}, \cite{lee2014spatial},\cite{musgrove2016fast}); while \cite{kim2010bayesian} proposed a mixture of experts model to represent spatial activation clusters. While these models have a number of different characteristics which make the approaches unique, most of them commonly assume a homogeneous, low order AR or ARMA process for the temporal noise. By homogeneous, we mean that the order of the AR or ARMA process is assumed constant across all voxels. This assumption is also made in \cite{penny2003variational}; however, as we demonstrate using a simple empirical example in the next section, this homogeneous AR order assumption may be violated in real fMRI data.

Instead of formulating the model at each voxel and then adopting spatial smoothing priors for parameters across voxels, another branch of research is based on vector autoregressive (VAR) processes, see \cite{harrison2003multivariate}. This approach allows for time-lagged dependence across voxels and spatial-temporal interaction but fitting these models across a large number of voxels is computationally intractable and low-rank approximations have to be used. These models are also useful for studying effective brain connectivity, where one time course is used to predict the other \citep{castruccio2016multi, chang2010time}. Another line of work chooses to model the temporal noise as a $1/f$ long memory process, and applies discrete wavelet transforms (DWT) towards fitting the model (see, e.g., \cite{jeong2013wavelet,bullmore2004wavelets,fadili2002wavelet,meyer2003wavelet}). While this approach seems promising, our focus in this paper will be with modeling short term memory using the classical AR process and spatial priors. The reason we choose to work with the AR process is because of its mathematical amenability and simplicity, and its wide use in different areas of science. A novel aspect of our work is that we allow the data to determine the order of the AR process at each voxel, borrowing strength from neighboring voxels, using ideas from Bayesian spatial variable selection.

Computation is an important issue when considering Bayesian spatial-temporal models for fMRI data. While the main focus of this paper lies with the development of a new model, another aspect of this work is the comparison of fully Bayesian and approximate Bayesian computation methods. Due to the computational burden associated with fitting models to high-dimensional brain imaging data, approximate Bayesian methods have received considerable attention in the neuroimaging literature. One such method is the variational Bayes (VB) inference \citep{penny2003variational,penny2007bayesian,woolrich2004constrained}. As there are currently no theoretical results quantifying the accuracy of VB methods (in contrast to MCMC which is justified by large sample theory of stationary Markov chains), the evaluation of VB has to be performed on a case-by-case basis. In some cases, the performance of VB can be quite good and in other cases it can be quite poor. In addition to the implementation of our new model based on a suitably designed MCMC sampler, we also develop an MCMC algorithm to sample from the posterior of the original GLM-AR model.  We then compare our model with both the VB implementation of the GLM-AR model (using SPM code) and our MCMC implementation of the GLM-AR model. Our studies indicate that under a low signal-to-noise (SNR) ratio the accuracy of MCMC outperforms VB according to several criteria.  %This finding can be considered an extension to our previous work \cite{teng2016comparison}. 

\subsection{Motivating example}\label{motexamp}
\begin{figure}[!ht]
\centering
\includegraphics[scale=0.75]{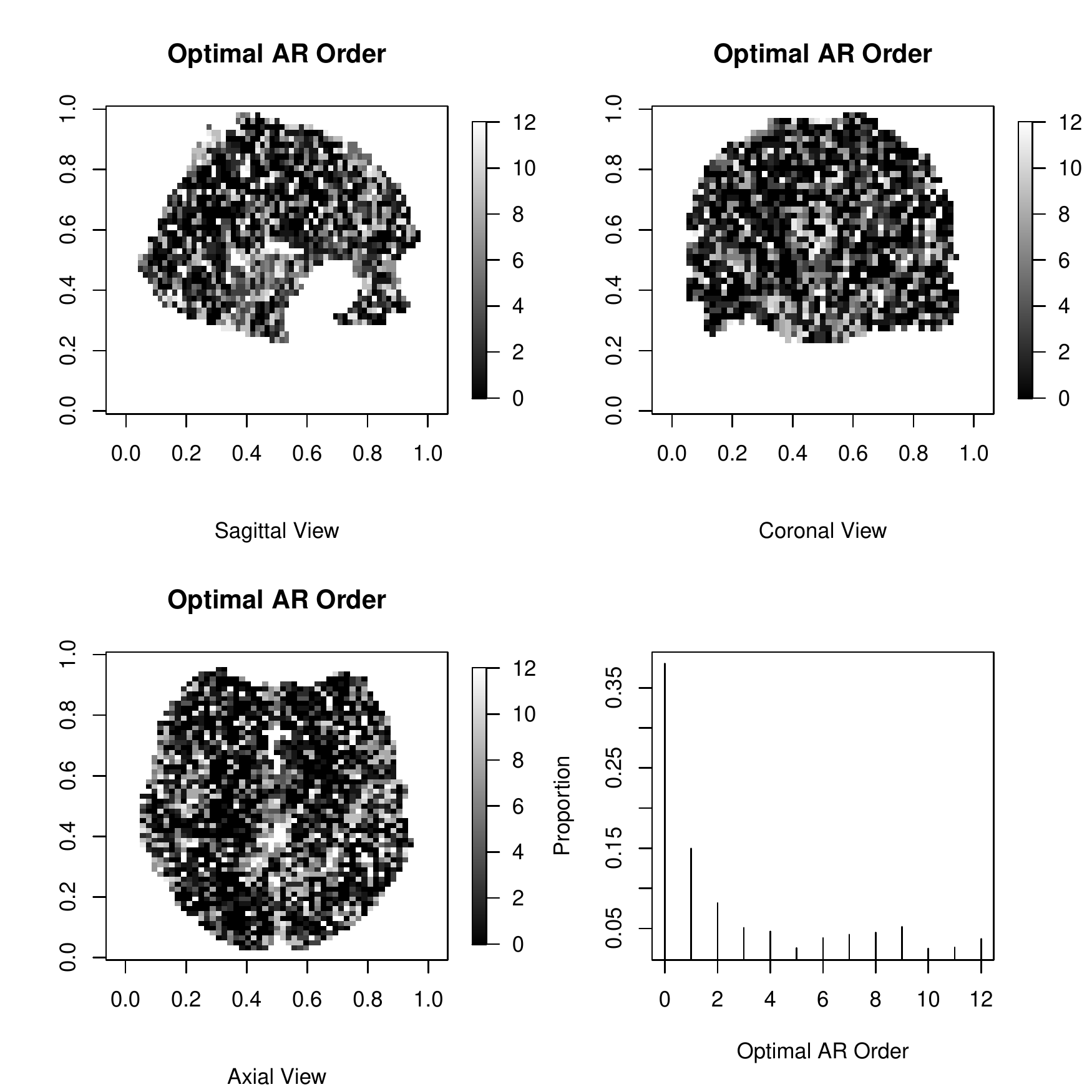}
\caption{\label{fig_OLS}Optimal maximum AR orders selected based on AIC. The upper AR order bound is set to 12.}
\end{figure}

Our motivating example comes from a single subject in a fMRI experiment examining a face-repetition stimulus. The experiment involves the presentation of either famous faces (F) or non-famous faces (N) with each type of face presented two times. Convolving this experiment design with the canonical hemodynamic response function and its time and dispersion derivatives leads to a design matrix with twelve columns plus one extra column for an intercept term in the regression model. After performing the necessary pre-processing steps as described in \cite{penny2005bayesian}, we fit a simple linear regression at each of the voxels. After obtaining the residuals from each voxel-specific fit, we fit an AR process up to order $12$ for each voxel using the ``ar'' function in R. We then selected the optimal AR orders based on the AIC criterion. Figure \ref{fig_OLS} displays a pictoral representation of the results.

Figure \ref{fig_OLS} shows considerable variablity in the estimated AR order across voxels. While most of the estimated optimal AR orders are 4 or less, higher orders up to $12$ are selected at some of the voxels. Furthermore, these estimated AR orders tend to show some extent of spatial clustering. If, as is often done, we simply model the data using a homogeneous low-order AR process, then the voxels with higher order estimated AR orders would be incorrectly modeled, and this inaccuracy in the modeling of temporal noise will have an impact on the inference on the covariates of interest (via underestimated standard errors), resulting in potentially false inferences about brain activation. To address this issue, we propose a spatially varying autoregressive order (SVARO) model, where the AR orders vary spatially across the brain. This is made possible by adopting a spike-and-slab prior with a stochastic search variable selection scheme. Spatial clustering of AR orders is incorporated by imposing an Ising prior \citep{ising1925beitrag} as the latent indicator for the spike and slab prior. We update the latent indicators using the Swendsen-Wang algorithm \citep{swendsen1987nonuniversal} alternating with Gibbs sampling in our MCMC algorithm. To prevent the phase transition problem associated with the Ising model, we derive theoretical bounds as in \cite{li2015spatial} and use these bounds to prevent critical slowing of the algorithm. We compare our model with the GLM-AR model of \cite{penny2007bayesian} (implemented under two schemes: our self written MCMC sampler and the VB algorithm available in the SPM software) in terms of mean squared error (MSE) and sensitivity. We conduct these comparisons using two simulation studies and then compare results on the motivating data set.

The rest of the paper is organized as follows: In Section \ref{methods} we present our model and MCMC sampling scheme. We present results from our simulation studies in Section \ref{sims}, followed by the analysis of the face-repitition data set in  Section \ref{app}. Lastly, we provide a discussion and outline some possible directions for future work in Section \ref{dis}.

\section{The Spatially-Varying Autoregressive Order Model}\label{methods}

First, we will define the model likelihood and then specify the spatial and temporal priors. Next, we discuss our posterior sampling scheme, include the construction of bounds for the hyperpriors that we use for the Ising prior on the AR orders.  Last, we discuss inference based on posterior probability maps.

\subsection{Model likelihood}

We let $P$ denote the maximum possible order of the AR process at each voxel and let $K$ denote the number of regression coefficients at each voxel. Using similar notation as in \cite{teng2016comparison}, for voxel $n \ (n=1,...N)$, we let $\mathbf y_n$ denote the observed time series of length $T$. For simplicity, our model is specified conditional on the first $P$ observations at each voxel so that the likelihood function is constructed based on the model for the remaining $T-P$ observations in the time series. We let $\mathbf X$ denote the $(T-P) \times K$ design matrix, $\mathbf w_n$ denote the $K$-dimensional vector of regression coefficients at voxel $n$, and $\mathbf e_n$ denotes the corresponding error term. Define the vector $\mathbf y_n \equiv \mathbf y_{1:T, n}$, the entire time series observed at voxel $n$.  The hierarchical model is specified in several stages. The first stage is a general linear model:
\begin{eqnarray}
\mathbf y_{P+1:T,n}  = \mathbf X \mathbf w_n + \mathbf e_n, 
\end{eqnarray}
where we emphasize again the implicit conditioning on $y_{1:P, n} \ (n=1,...N)$. Let $\tilde{\mathbf E}_n$ denote the embedded error (or lagged prediction) matrix of dimension $(T-P) \times P$, with $t,p$ element $(\mathbf y_{P+1:T,n} - \mathbf X \mathbf w_n)_{[t-p]}$ where the notation $[i]$ denotes the $i^{th}$ index of the vector. Let $\mathbf z_n \equiv z_{P+1:T,n}$ denotes a vector of i.i.d mean-zero Gaussian random variables with precision $\lambda_n$. The second stage is then an AR model at each voxel:
\begin{eqnarray}
\mathbf e_n = \tilde{\mathbf E}_n \mathbf a_n + \mathbf z_n 
\end{eqnarray}
where $\mathbf a_n$ is a vector of autoregressive coefficients for the time series at voxel $n$.

Letting $c$ denote a constant term, the log-likelihood for voxel $n$, is
\begin{equation} 
l_n=- \frac{\lambda_n}{2} \sum_{t=P+1}^T \left [  (y_{tn}- \mathbf x_t \mathbf w_n)- \mathbf{\tilde e}_{tn} \mathbf a_n    \right ]^2 + \frac{T-P}{2} \log \lambda_n + c,
\end{equation}
where $\mathbf x_t$ is the $(t-P)$th row of the design matrix $\mathbf X$ and $\mathbf{\tilde e}_{tn}$ is the $(t-P)$th row of $\tilde{\mathbf E}_n$.

Summing this log-likelihood over the number of voxels, $n$, we obtain the overall likelihood:
\begin{equation}
l=\sum_{n=1}^N \left \{ - \frac{\lambda_n}{2} \sum_{t=P+1}^T \left [  (y_{tn}- \mathbf x_t \mathbf w_n)- \mathbf{\tilde e}_{tn} \mathbf a_n    \right ]^2 + \frac{T-P}{2} \log \lambda_n + c. \right \}
\end{equation}

\subsection{Spatial prior}\label{sub_spatial}
At the next level of the model we specify a spatial smoothing prior for the regression coefficients $\mathbf W=(\mathbf w_1, ..., \mathbf w_N)$, a $K \times N$ matrix, with $k$th row $\mb W_{k,.}$. Following \cite{penny2005bayesian}, we assume that the prior for $\mathbf W$ takes the form
\begin{eqnarray} 
 \pi(\mathbf W)&=&\prod_{k=1}^K \pi(\mb W_{k,.}) \label{eq_W1} \\
 \mathbf W_{k,.} &\sim& \mbox N\left(\mathbf 0,\alpha_k^{-1} (\mathbf S^T \mathbf S)^{-1}\right),\quad k = 1,\dots,K. \label{eq_W2}
\end{eqnarray}
A priori the regression coefficients within voxels are independent (\ref{eq_W1}) whilst spatially, the $k$th, $k = 1,\dots,K$, regression coefficients (across voxels) are modeled dependently through an $N$-dimensional multivariate normal distribution (\ref{eq_W2}).
Here $\mathbf S$ is known as a Laplacian matrix. The $n$th diagonal term of this matrix is equal to the corresponding number of first order neighbors of the voxel $n$.  All off-diagonal terms are zero except for $-1$ in off-diagonal elements $(n,j)$ and $(j,n)$ if voxel $j$ is a neighbor of voxel $n$, for $n = 1,\dots,N$. This form for the prior accommodates spatial smoothing while also being sparse and convenient to work with computationally. In the SPM12 software this prior is referred to as the ``LORETA'' prior. Ultimately, what is of primary interest in studies of brain activation is a posterior probability of some function of these regression coefficients, and this posterior probability is computed at each voxel to produce posterior probability maps (PPM, see \cite{penny2005bayesian}). The precision parameter in (\ref{eq_W2}), $\alpha_k$, is assigned a conditionally conjugate hyper-prior:
\begin{equation}
\alpha_k \stackrel{iid}{\sim} \mbox{Gamma} (q_1, q_2) \ (k=1,...K).
\end{equation}

\subsection{Temporal prior}
The key difference between our model and the model of \cite{penny2007bayesian} lies in our modeling of the temporal noise. Rather than assuming AR orders are homogeneous across the brain (we refer the readers to \cite{teng2016comparison} and \cite{penny2007bayesian} for model details), we allow for variability in the order of the AR processes across voxels. In addition, we adopt a spatial prior for this variability under the assumption that the AR orders of neighboring voxels will be similar. Specifically, for each voxel $n$ and order $p$, $p=1,\dots,P$, we assign the latent indicator variable $\gamma_{pn}$ to the $p^{th}$ AR coefficient $a_{pn}$, such that given $\gamma_{pn}\ (p=1...P, n=1...N)$, $a_{pn}$ will be conditionally independent. $\gamma_{pn}$ will take value $1$ if order $p$ is present for voxel $n$ and $0$ otherwise. Conditional on $\gamma_{pn}$, $a_{pn}$ will either have a normal distribution or unit mass at $0$. This is commonly referred to as the spike-and-slab prior  \citep{george1993variable,mitchell1988bayesian}, though we note that our formulation is a \emph{spatial} spike-and-slab prior and that this prior is assigned to the coefficients of the AR process governing the temporal noise.
\begin{eqnarray*}
  \pi(\mathbf a \mid \boldsymbol\gamma) &=& \prod_n \prod_p \pi(\mathbf a_{pn} \mid \gamma_{pn}) \\
  \pi(a_{pn} \mid \gamma_{pn}) &=& \gamma_{pn} \phi (a_{pn}; 0, \tau_p^2) + (1-\gamma_{pn}) I_0(a_{pn}) 
\end{eqnarray*}

Here, $\phi(\cdot;a,b)$ is the pdf of a normal distribution with mean $a$ and variance $b$ and $I_0(\cdot)$ is the indicator function that its argument equals 0, and where $\gamma_{pn}$ is the binary indicator. $\tau_p$ is the precision of the normal component and is again given a Gamma prior $\tau_p \sim \mbox{Gamma} (r_1, r_2)$. 

The advantages of introducing such a prior are three fold: First, the orders in the AR process at each voxel that lack support from the data can be effectively removed from the model as the corresponding AR coefficients can be shrunk exactly to $0$. This allows us to infer which orders are present in which voxels. Second, the number of voxels with high AR orders is non-zero but expected to be small, which is an aspect of this prior that can be controlled
by tuning the hyper-parameters. Third, for some of the voxels there might be vacancies in some of the middle orders while there are some non-zero coefficients for higher orders. The proposed model is flexible enough to allow for this behaviour, since we have a total of $P$ independent Ising processes, one for each possible order $p \in \{1,\dots,P\}$. 

There are of course other model selection techniques that could have been considered. For example a type of Bayesian lasso could have bee used as an alternative to the spike-and-slab prior. \cite{wang2007regression} have applied the lasso to the selection of AR processes, and for Bayesian lasso we refer to \cite{schmidt2013estimation}. A recent alternative prior known as the "non-local" prior for variable selection has been proposed by \cite{johnson2012bayesian} and has been demonstrated to have desirable consistency properties and yield smaller prediction errors in large sample settings. A review of Bayesian priors that can be employed for model selection is presented in \cite{o2009review}. 

We assume that the indicator processes are independent across different orders, $\pi(\bs \gamma)=\prod_p \bs \gamma_p$, where $\bs \gamma_p=(\gamma_{p1},...,\gamma_{pN})^T$. The simplest variable selection model would assume $\gamma_{pn}$ follows a Bernouli distribution \citep{george1993variable}. Here, in order to borrow information across neighbors as well as to model the spatial clustering effect of AR orders, we choose to use the Ising prior \citep{smith2007spatial} independently for each $p=1...P$. 
\begin{eqnarray}
P(\bs \gamma_p)\propto \exp \left (\beta_{0p} \sum_n \gamma_{pn} + \beta_{1p} \sum_{n_1 \sim n_2} I( \gamma_{p n_1}= \gamma_{p n_2}) \right ),
\end{eqnarray}
where $\beta_{0p}$ and $\beta_{1p}$ are two hyper-parameters controlling the sparsity and smoothness of the binary latent field, respectively. Typically, a higher value of $\beta_{0p}$ results in less sparsity and a higher value of $\beta_{1p}$ indicates more smoothness. One issue with the Ising model that requires some care is the choice of hyper-parameters. When these parameters take values near what is known as the ``phase transition'' boundary, the mixing of an MCMC sampler will suffer from critical slow down \citep{stanley1987dynamics}. To avoid the phase transition boundary,  we adopt an analytical approach similar to \cite{li2015spatial} to quantify the value for the bounds for $\beta_{0p}$ and $\beta_{1p}$. An outline of the bound derivation is given in Subsection \ref{sec_Ising}. %Details are given in {\color{red} REFERENCE MING's DISSERTATION}.

\subsection{Posterior sampling scheme}
Most parameter updates related to the posterior sampling of our model can be accomplished via Gibbs sampling. 
%{\color{red} REFERENCE MING's DISSERTATION}. 
One exception is the update to the latent indicator $\bs \gamma$. For $\bs \gamma$ we use the Swendsen-Wang algorithm alternating with Gibbs updates \citep{johnson2013bayesian}. This strategy, that is, mixing Swendsen-Wang updates with Gibbs updates for $\bs \gamma$ has proven successful in improving the mixing of the Markov chain sampler and results in faster block updates in various studies \citep{higdon1998auxiliary}.

\subsection{Bound construction for the Ising Model Hyper-Parameters}\label{sec_Ising}
The hyper-parameters in the Ising prior play a vital role in posterior estimation. Without careful selection, we are faced with mixing issues associated with ``phase-transition" \citep{stanley1987dynamics}. There are various approaches to sampling such hyper-parameters, \cite{johnson2013bayesian} estimated them using path sampling \citep{gelman1998simulating}, \cite{shu2015multiple} proposed a Monte Carlo EM algorithm to obtain a point estimate of the hyper-parameters, but these procedures would be too time consuming for our model, considering that we have over $10$ independent Ising fields and thousands of voxels. \cite{smith2007spatial} proposed to update the hyper-parameters and binary indicators together, but this approach still suffers from potential possibility of sampling over the phase transition boundary. Here, we adopt an approach similar to that considered in \cite{li2015spatial} and construct some theoretical bounds to prevent phase transition. The resulting hyper-parameter values are then chosen as fixed values within the estimated bounds. This procedure turns out to work well in our analysis and studies. 

To construct the bounds, we first write out the posterior conditional density with respect to $\bs \gamma_p$, 
\begin{eqnarray}
P(\bs \gamma_p \mid \cdot) \propto && \exp \Big (   \beta_{0p}\sum_{n}\gamma_{pn} +\beta_{1p} \sum_{n_1 \propto n_2} \mbox I \{ \gamma_{pn_1}=\gamma_{pn_2}    \}   \\
&&+  \sum_n \sum_t \frac{-\lambda_n}{2} (e_{tn}-\sum_{p} \tilde e_{tnp} a_{pn})^2  \Big ) \nonumber
\end{eqnarray}
In our model where multiple orders exists across space, it is natural to assume that: 1) there are relatively few time series with large AR order; and 2) the posterior density when low AR orders exist is greater than that when low AR orders do not exist, meaning $P(\bs \gamma_p \mid \cdot)$ is greater than $P(\mathbf 0 \mid \cdot)$.  

Let $\pi_p$ denote the candidate voxels selected for order $p$. The maximum number of first order neighbors is 8. Let $V_p=(\pi_p  N)^{1/3}$ denote the length of an edge a voxel, then there are $3V_p^2(V_p-1)$ neighboring pairs. Based on this we derive 
\begin{eqnarray}\label{eq_V}
\beta_{0p}\sum_n \gamma_{pn} +\beta_{1p} \sum_{n_1 \sim n_2} \mbox I(\gamma_{pn_1}=\gamma_{pn_2}) = \beta_{0p}V_p^3+3\beta_{1p}V_p^2(V_p-1) 
\end{eqnarray}
According to 1), we know that for high AR orders (typically $P>8$) $\beta_{0p}+3\beta_{1p}<0$. According to 2), we know that for low AR orders (typically $P<4$), 
\begin{eqnarray}
&&\sum_n \sum_t \frac{-\lambda_n}{2} (e_{tn}-\sum_{p_0 \neq p}\tilde e_{tnp_0}a_{p_0n})^2 \leq \sum_n \sum_t \frac{-\lambda_n}{2} (e_{tn} \\
&&-\sum_{p_0 }\tilde e_{tnp_0}a_{p_0n})^2  +\Big[  \beta_{0p} \sum_n \gamma_{pn} +\beta_{1p} \mbox I(\gamma_{pn_1}=\gamma_{pn_2})  \Big] \nonumber
\end{eqnarray} 
Reorganizing this by moving the first term on right-hand side to left produces:
\begin{eqnarray}
&&\sum_n \sum_t  \frac{-\lambda_n}{2} \left[ (e_{tn}-\sum_{p_0 \neq p}\tilde e_{tnp_0}a_{p_0n})^2 - (e_{tn} -\sum_{p_0 }\tilde e_{tnp_0}a_{p_0n})^2  \right]\\
&& \leq \Big[  \beta_{0p} \sum_n \gamma_{pn} +\beta_{1p} \mbox I(\gamma_{pn_1}=\gamma_{pn_2})  \Big] \nonumber
\end{eqnarray}
The two terms in the bracket on the left side can be considered as one with and without $\tilde e_{tnp}a_{pn}$. Thus, it can be roughly considered as the residual sum of squares of a common linear regression when $a_{pn}$ is included in the model or not. Let $R_{pn}^2$ denote the coefficient of determination for voxel $n$ and order $p$, then we have
\begin{eqnarray}
\beta_{0p}\sum_n \gamma_{pn} +\beta_{1p}\sum_{n_1 \sim n_2} \mbox (\gamma_{pn_1} =\gamma_{pn_2}) \geq -\frac{1}{2} \sum_n \sum_t \frac{R_{pn}^2}{1-R_{pn}^2}
\end{eqnarray}

Combined with Equation \ref{eq_V}, we have
\begin{eqnarray}
\beta_{0p}V_p^3 +3\beta_{1p} V_p^2(V_p-1) \geq -\frac{1}{2}\pi_p N T   \frac{R_{pn}^2}{1-R_{pn}^2}
\end{eqnarray}
For a 3-dimensional grid we assume $N=56526$ as the number of voxels. Among them, a proportion of $\pi_p=0.1$ are selected for order $p$. So $V_p=(\pi_p N)^{1/3}=17.8$. We assume that 5\% of the variation can be explained as a result of order $p$, so $R_{pn}^2=0.05$. We then have $\beta_{0p}+2.83\beta_{1p} \geq -9.26$.  

Note that the inequality above just gives a range values for the hyperparameters, rather than providing the values directly. In practice, the exact values of hyperparameters are largely determined by the researcher, which should be combined with one's prior experience and an initial analysis of the data. We suggest obtaining such values based on some exploratory ad-hoc approaches, e.g., a linear regression at each voxel followed by fitting an AR process. Then the the estimated optimal orders can be used as a reference when determining the hyper-parameters in the Ising model. This method has turned out to work well empirically as we demonstrate in Section 5.

\subsection{Posterior probability maps}
A primary emphasis on fMRI data analysis is inference for activation. For Bayesian modeling of fMRI data this is typically achieved via the PPM. % so we provide background on posterior probability maps (PPM).% A contrast for a certain voxel $n$ is the inner product of a contrast vector $\mathbf c$ with the regression coefficient in that voxel $\mathbf w_n$. The contrast vector $\mathbf c$ is typically a weighted vector with elements consisting of $1$ and $-1$ representing an effect of interest. For example, to study the effect of A versus B when these are the only two conditions would lead to a contrast vector of $\mathbf c=(1,-1)^T$. 
Let $\mathbf c^T\mathbf w_n$ denote a contrast of interest of the regression coefficients.
A PPM is a map of the posterior probability of activation for each voxel: $\Pr(\mathbf c^T \mathbf w_n > \delta_e \mid \mathbf y)$. Here $\delta_e$ is a pre-specified ``activation threshold", for example, a value that corresponds to $1\%$ of the global mean value. Thus, PPM looks at the probability of the contrast $\mathbf c^T \mathbf w_n$ being greater than activation threshold $\delta_e$, given the data.  

To formally determine activation in the brain, one can look at a thresholded PPM. This is obtained by exerting a second threshold, namely a ``probability threshold" $\delta_p$, onto the original PPM. Thus, a voxel is ``activated" if $\Pr(\mathbf c^T \mathbf w_n >\delta_e \mid \mathbf y)>\delta_p$. This $\delta_p$ reflects the confidence of the inference and usually takes a value above $0.9$ (e.g. $0.95$ or $0.99$). This process discretizes the PPM into ``null" and "activated" voxels and is commonly used in summarizing a Bayesian analysis for brain activation.

\section{Simulation Study}\label{sims}

To evaluate the performance of our model, we make comparisons with the standard GLM-AR spatial model. One implementation of this model that we make comparisons to is the Variational Bayes (VB) method available in SPM12 software.  Another implementation is our self-written MCMC sampler for the same model. Although the accuracy of VB has been verified in a setting with high signal-to-noise ratio (SNR) by \cite{teng2016comparison}, under low SNR, MCMC outperforms VB according to certain metrics. This will be illustrated presently. Henceforth, we will refer to the VB implementation and MCMC implementation of the GLM-AR model as PVB and PMCMC, respectively.

\subsection{Simulation design}
%The synthetic data is modeled after face-repetition data sest\citep{henson2002face} for a single subject. Complete information on the data set can be found online at \\
%\emph{http://www.fil.ion.ucl.ac.uk/spm/data/}. In this experiment, famous faces and non-famous faces are presented two times, resulting in four types of stimuli (F1,F2, U1, U2) where F denotes a famous face, U denotes an unfamiliar face, whereas 1 and 2 refer whether the face is presented the first or second time. These stimuli are convolved with the canonical HRF to be formally used as regressors in the statistical model. In terms of voxels, we perform our simulations on a 2-dimensional axial slice of the brain consisting of $53\times 63 \times 52$. 

Our design matrix consists of two columns ($K=2$); the first column is the experimental design (fashioned after the face-repetition data set) convolved with the canonical HRF, and the second column is the intercept (a vector of $1$s). The parameters of interest, corresponding to the experimental design, (one at each voxel) are generated under a mean zero multivariate normal distribution: $\mathbf W_{1,.} \sim N(0,(10\mathbf S^T \mathbf S)^{-1/2})$, while the intercepts are generated under a mean $100$ multivariate normal distribution: $\mathbf W_{2,.} \sim N(100,(10\mathbf S^T \mathbf S^T)^{-1/2})$. The model noise will have a precision of $\lambda_n=0.1 \ (n=1...N)$. This corresponds to a fairly small signal-to-noise (SNR) ratio, where the temporal noise will play a greater role in the data. In the following, we will carry out two simulations. In the first case the data will be simulated under our model, and in the second case the data will be generated under the standard spatial GLM-AR model. In these two simulations, we investigate the estimation accuracy of the slopes ($\mathbf W_{1,dot}$), intercepts ($\mathbf W_{2,dot}$), and autoregressive coefficients ($\mathbf a_p \ p=1...P$), and we also examine if the difference in inference for these coefficients will lead to a possible difference in the final inference on brain activation. All simulations are based on $100$ replicate data sets, and 
we perform the simulations on a 2-dimensional axial slice of the brain.

\subsection{Simulation 1}
Here we simulate the AR parameters under our SVARO model, that is from the Ising prior. We assume that the maximum order at each voxel is $P=8$. The precision parameters are set as $\tau_p=20 \ (p=1...P)$. For simplicity, we assume that all AR orders are generated spatially according to the same values for the hyperparameters of the Ising model, i.e., $\beta_{0p}=-0.2$ and $\beta_{1p}=0.3$. The AR order in PMCMC and PVB are set to $P=1$ as is fairly standard practice. We note here that the GLM-AR model is misspecified and we expect that its performance will suffer regardless of which posterior sampling algorithm is used.

\begin{figure}[!ht]
\centering
\includegraphics[scale=0.4]{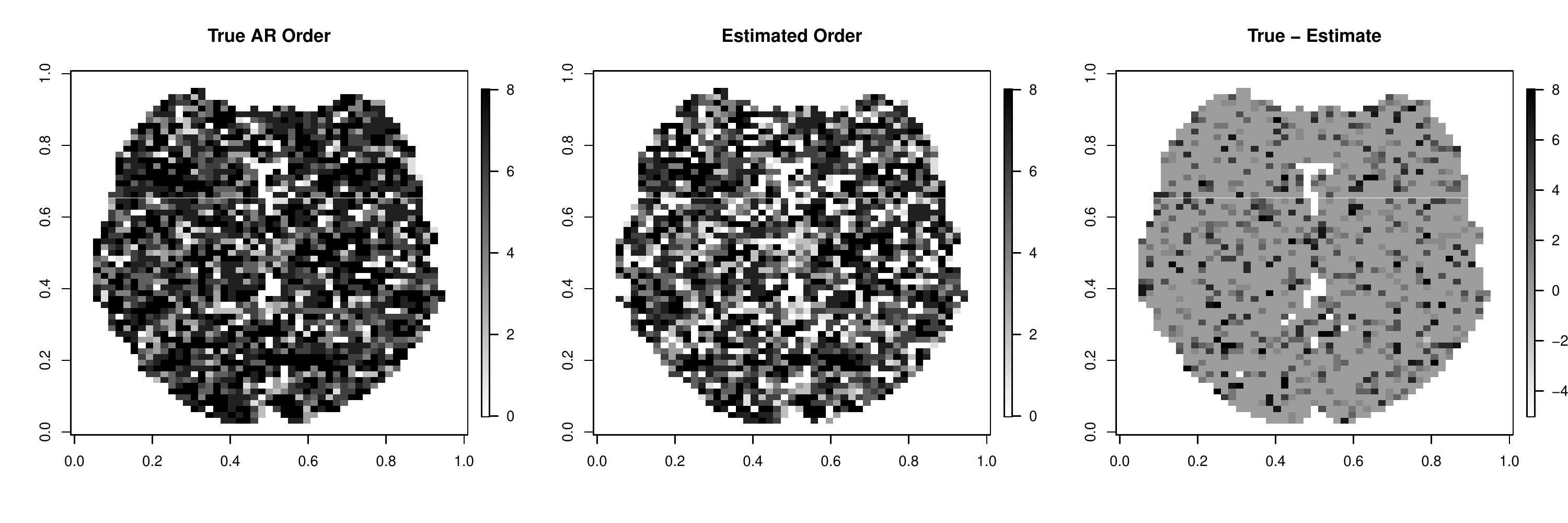}
\caption{\label{fig_orders} Maximum orders of AR coefficients in each voxel. The left figure displays the truth, ranging from 0 to 8. The middle figure shows the posterior estimates of the maximum orders from our SVARO model. The right figure displays the difference between the two.}
\end{figure}

Figure \ref{fig_orders} shows the true AR orders, the estimated maximum orders from our SVARO model, and the difference of the two. The estimated maximum orders are obtained by averaging the posterior mean at each voxel over their simulation replicates and rounding. We can see that most of the orders match between the two figures indicating good performance. However, there are some negative values in the difference map. 

\begin{figure}
\centering
\includegraphics[scale=0.5]{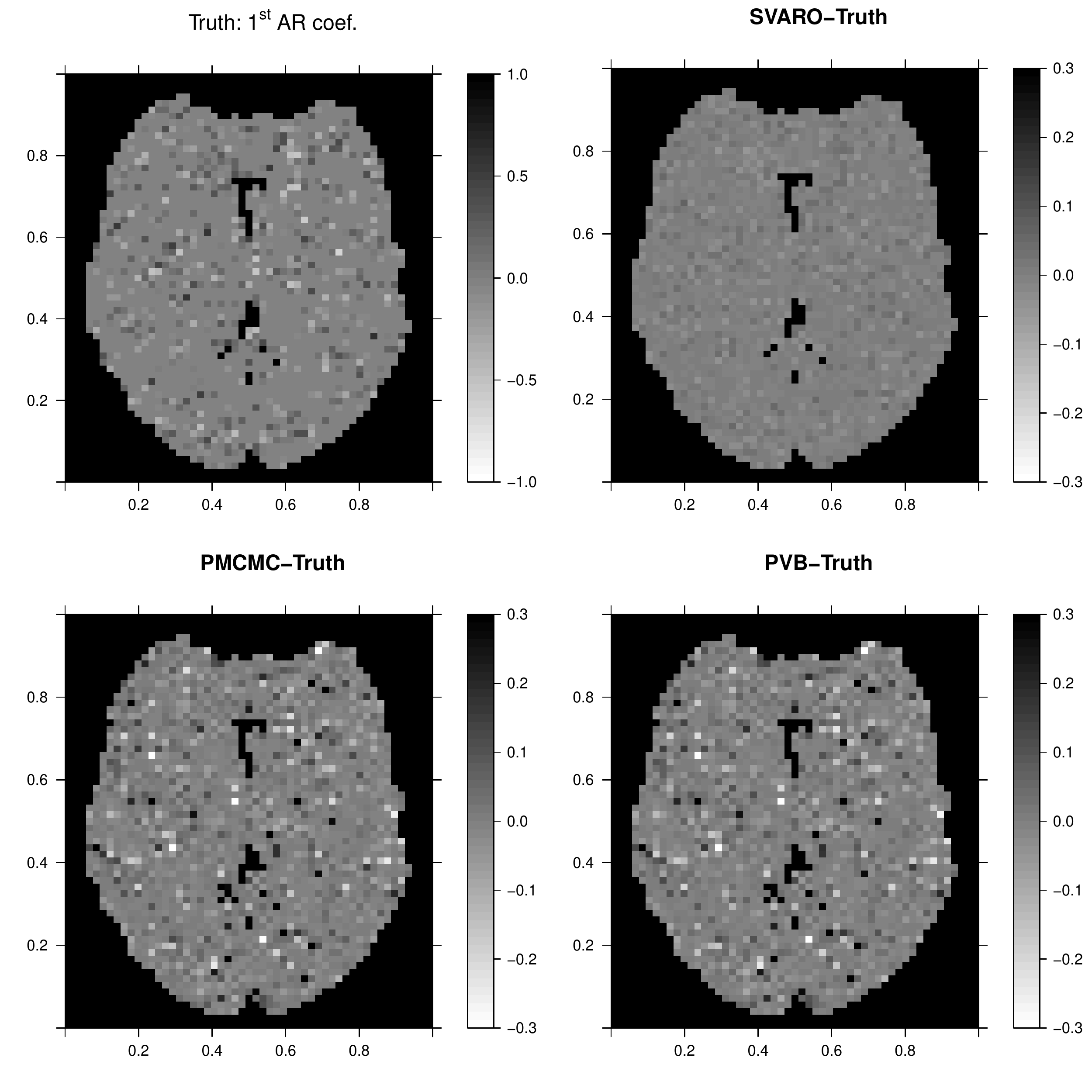}
\caption{\label{fig_AR}The top left figure displays the true AR coefficients of order one. The remaining figures display difference maps between the true order one AR coefficients and the estimated posterior mean from SVARO, PMCMC, and PVB. The posterior means are averaged over the 100 replicated simulation data sets.} 
\end{figure}

Next we compare SVARO with PVB and PMCMC in estimating the $1$st AR coefficient. As shown in Figure \ref{fig_AR}, SVARO shows little error compared with the truth, indicating that our model has captured the autoregressive parameter quite well. In contrast, PMCMC and PVB exhibit more bias, indicating a lack of fit for the temporal noise. Note that we are only displaying the SVARO estimates for the $1$st AR coefficient for simplicity and direct comparison, the other AR coefficients are similarly well-estimated.

Table \ref{tab_MSE} summarizes the average MSE for various parameters. These summaries are obtained by averaging the MSE of the corresponding parameters across all the voxels and over simulation replicates. It is clear that SVARO has the smallest MSE for these three parameters. In addition, PMCMC outperforms PVB in estimating the coefficient of the hemodynamic response, which is the primary parameter upon which inference is based. This finding is in line with our previous findings in \cite{teng2016comparison}, where we found that a low SNR is one setting where MCMC outperforms VB for this particular model (GLM-AR). We also calculate and present the log-pseudo marginal likelihood (LPML, \cite{gelfand1994bayesian}) in Table \ref{tab_MSE}. SVARO has a larger LPML than the MCMC implementation of GLM-AR, indicating a better model fit. Note that LPML cannot be obtained from the VB algorithm. In terms of timing, SVARO takes 108 minutes with 10,000 iterations following 10,000 burn-in iterations, PMCMC takes 11 minutes with the same number of iterations, PVB is the fastest at 1 minute computation time. 

\begin{figure}[!ht]
\centering
\includegraphics[scale=0.45]{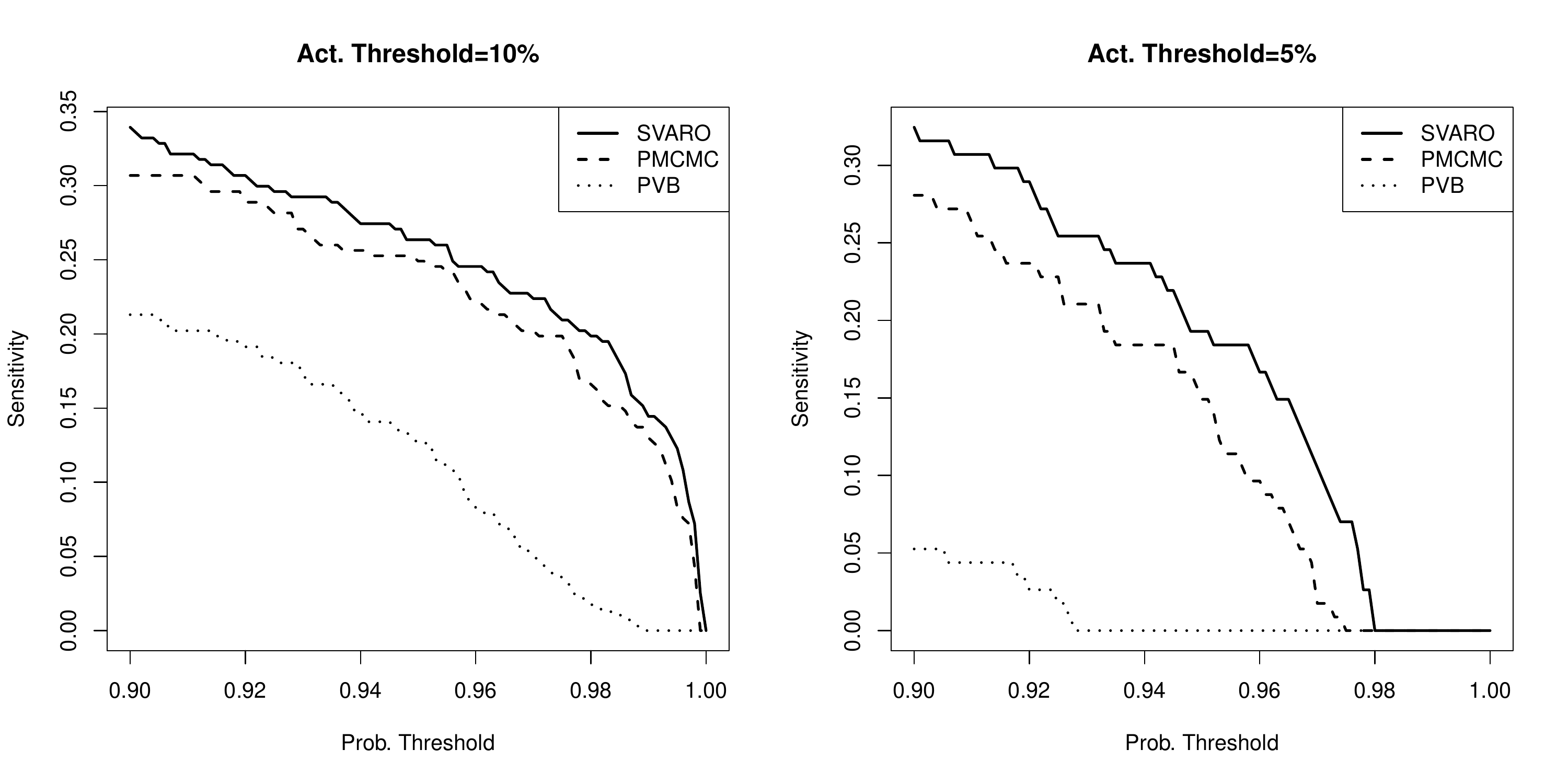}
\caption{\label{fig_sensitivity} Thresholded sensitivity curves for the three methods, with two activation thresholds. Left: acitvation threshold corresponds to the top 10\% of the parameter estimates; Right: top 5\%. The x-axis denotes the probability threshold values and y-axis denotes the corresponding sensitivity.}
\end{figure}

We next investigate how the differences observed for the individual parameters will impact the overall inference of interest. A sensitivity plot is presented in Figure \ref{fig_sensitivity}. This figure is obtained by plotting the average sensitivity against a range of marginal posterior probability thresholds from $0.9$ to $1$. We choose this range because it covers those values most often used in practice.
\begin{table}
    \caption{\label{tab_MSE}Table of MSE, LPML and Timing for the three models. MSE is calculated by averaging MSE in each voxel and over simulation replicates. The MSE values for PMCMC and PVB are relative to those in SVARO.}
    
    \centering
    \begin{tabular}{lrrrrr}\hline\hline
        	&		&	MSE	&		&	LPML	&	Timing	\\ \cline{2-4}
	&	$\mathbf W_{1,.}$	&	$\mathbf W_{2,.}$	&	$\mathbf {a_1}$	&		& (min) \\ \hline
	SVARO	&	0.478	&	0.030	&	0.001	&	-1842902	&	108	\\
	PMCMC	&	113\%	&	135\%	&	509\%	&	-1926620	&	11	\\
	PVB	&	199\%	&	138\%	&	510\%	&	NA	&	1	\\ \hline\hline
    \end{tabular}
\end{table}

In terms of the underlying activation threshold, we use two thresholds: the true value of the contrast that corresponds to the top $10\%$ and top $5\%$ of all voxels. Thus, corresponding to a certain activation threshold and a certain probability threshold, the higher the sensitivity, the better the model is in terms of capturing activation. Again, a notable difference is observed when comparing the three methods, with SVARO giving the uniformly highest sensitivity across the entire range of probability thresholds and PVB resulting in the lowest sensitivity. PMCMC is better than PVB but still under-performs relative to SVARO.

\begin{figure}[!ht]
\centering
\includegraphics[scale=0.55]{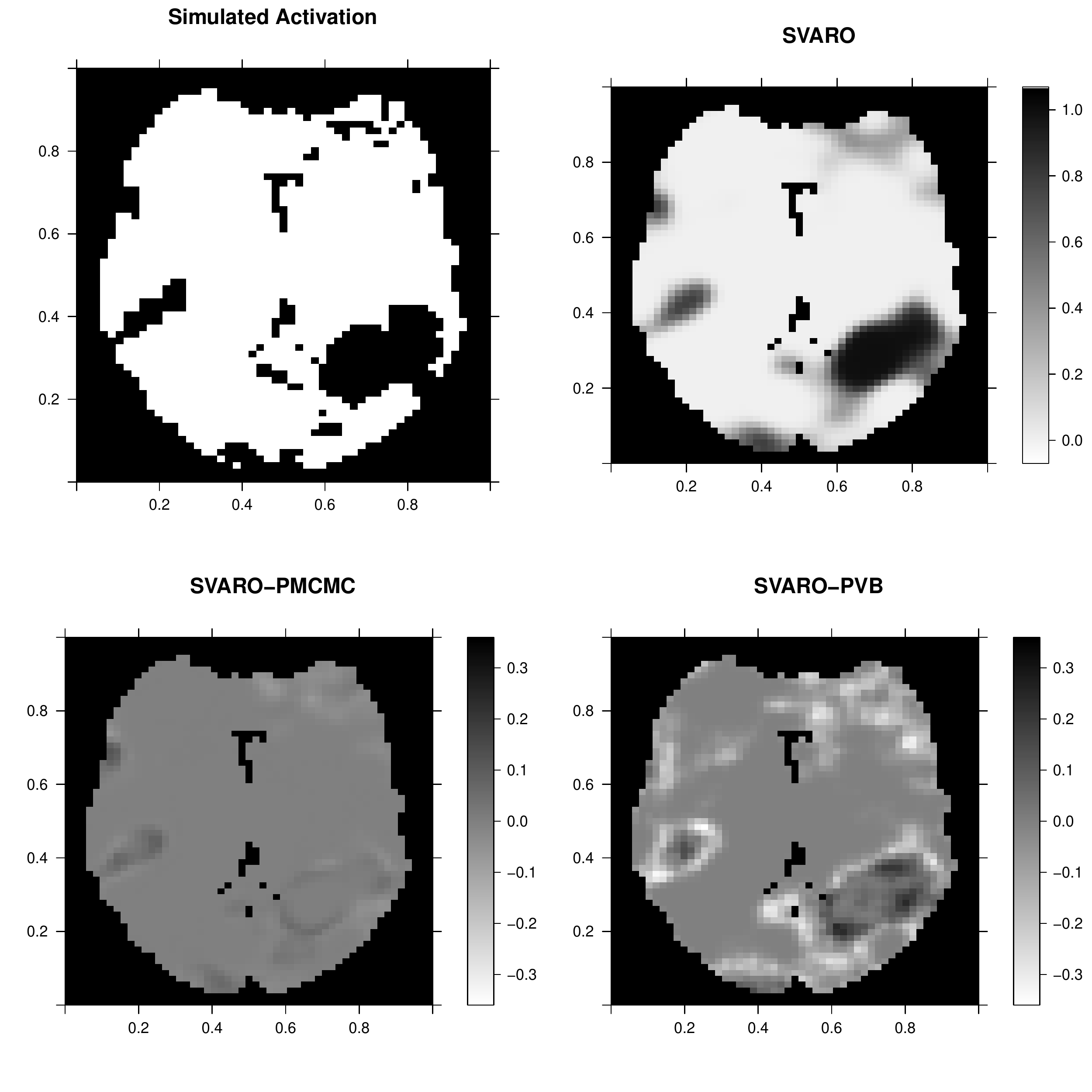}
\caption{\label{fig_ppm}Top left: the true activation map (red denotes activated voxels). The remaining panels are posterior probability maps (PPM) of activation obtained using SVARO, (SVARO-PMCMC) and (SVARO-PVB). The latter two reflect the difference of the two alternative approaches relative to SVARO.}
\end{figure}

We plot the posterior probability maps (PPM) in Figure \ref{fig_ppm}. This figure depicts the locations of the true activations and the posterior probability maps from SVARO. In addition, differences in the probability maps comparing SVARO with PMCMC and PVB are also depicted. Again, SVARO appears to perform the best in producing the highest posterior probabilities for regions that are truly activated. PMCMC is similar to SVARO but its probability on those activated regions are slightly lower than those from SVARO, especially on the boundary. PVB under performs compared with the other two approaches by providing greater posterior probability on null locations while providing smaller posterior probability on actived locations.

\subsection{Simulation 2}
Although the real data examined earlier suggest the existence of heterogeneous AR orders we want to compare the performance of SVARO under a homogeneous AR order assumption, where now the GLM-AR model is correctly specified. To do so, we simulate under the competing GLM-AR model. The AR coefficients are simulated under the LORETA prior and the AR order is set to $1$ for every voxel, with prior precision $\tau_p=400$. We set the maximum AR order to $P=12$ when applying SVARO. Thus, PMCMC and PVB are working under the true model while SVARO is working under a more general model. 

Table \ref{tab_MSE_contrary} shows the MSE summaries of the estimators. When data are simulated under the competing model, SVARO still exhibits good performance in terms of MSE for the two regression parameters. Its MSEs are only slightly larger the those under the correctly specified model using MCMC posterior simulation. It is worth mentioning that PVB again under performs relative to PMCMC in terms of the hemodynamic response parameter and the AR coefficient. 

\begin{table}
    \caption{\label{tab_MSE_contrary}MSE, LPML and computation time for the three methods. MSE is calculated by averaging MSE over voxels and simulation replicates. The MSE values for PMCMC and PVB are relative to those in SVARO.}

    \centering
    \begin{tabular}{lrrrrr}\hline\hline
 	    &		&	MSE	&		&	LPML	&	Timing	\\ \cline{2-4}
    	&	$\mathbf W_{1,.}$	&	$\mathbf W_{2,.}$	&	$\mathbf a_1$	&		& (min)		\\ \hline
        SVARO	&	0.502	&	0.031	&	0.003	&	-1817287	&	206	\\
        PMCMC	&	99\%	&	97\%	&	54\%	&	-1875900	&	11	\\
        PVB	&	167\%	&	98\%	&	49\%	&	NA	&	1	\\ \hline \hline
    \end{tabular}
\end{table}
%\begin{table}
%\caption{\label{tab_MSE_contrary}MSE, LPML and computation time for the three methods. MSE is calculated by averaging MSE over voxels and simulation replicates. The MSE values for PMCMC and PVB are relative to those in SVARO.}
%\centering
%\begin{tabular}{lrrrrr}
%\hline \hline
%	&		&	MSE	&		&	LPML	&	Timing	\\
%	&	$\mathbf W_{1,.}$	&	$\mathbf W_{2,.}$	&	$\mathbf a_1$	&		& (min)		\\ \hline
%SVARO	&	0.502	&	0.031	&	0.003	&	-1817287	&	206	\\
%PMCMC	&	99\%	&	97\%	&	54\%	&	-1875900	&	11	\\
%PVB	&	167\%	&	98\%	&	49\%	&	NA	&	1	\\				
%\hline \hline
%\end{tabular}
%\end{table}

\begin{figure}[!ht]
\centering
\includegraphics[scale=0.45]{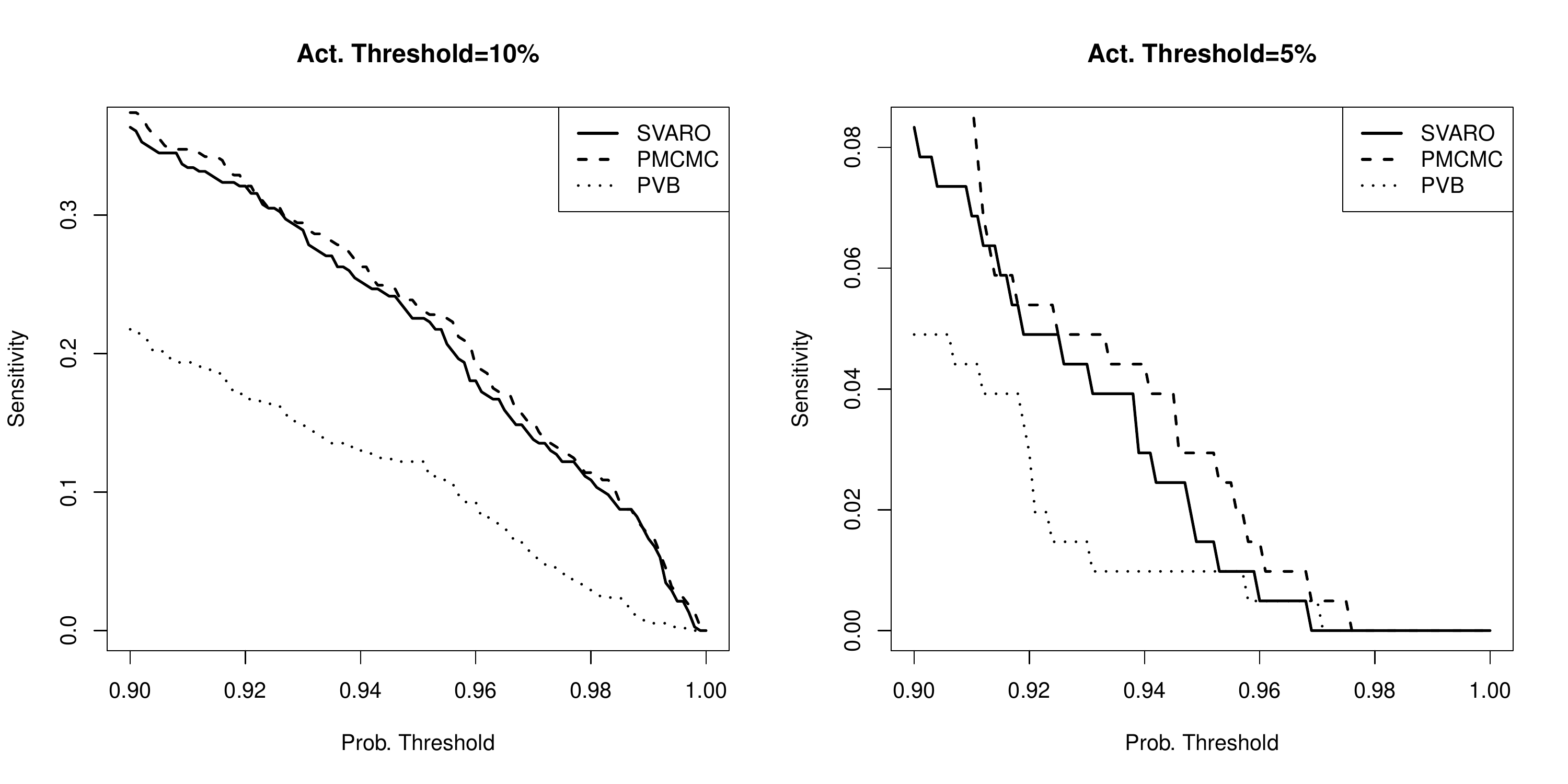}
\caption{\label{fig_sensitivity_contrary} Thresholded sensitivity curves for the three methods, with two activation thresholds. Left: activation threshold corresponds to the top 10\% of the parameter estimates; Right: top 5\%. The x-axis denotes the probability threshold values and y-axis denotes the corresponding sensitivity.}
\end{figure}

Figure \ref{fig_sensitivity_contrary} presents the sensitivity curves. Despite the data being simulated under a constant order AR assumption, SVARO demonstrates similar sensitivity to that of PMCMC. The sensitivity curve for PVB is uniformly smaller than the other two because of the inaccurate estimation of the hemodynamic response parameter.

Overall, these simulation studies show that our SVARO model is more efficient from a statistical point of view than the GLM-AR model when the AR orders of voxels differ---a situation we believe to be more common than not.  They also show that our model does not suffer much in terms of statistical efficiency when the AR order is constant across voxels.

\section{Analysis of Face Perception fMRI Data}\label{app}
We turn our focus back to the face repetition data set that originally motivated our model development and compare results from the two models and three algorithms. In this analysis we use the complete experimental design consisting of famous faces, repeated twice and unfamiliar faces, repeated twice. These four design vectors are then convolved with the canonical HRF as well as its time and dispersion derivatives.  An intercept term is also added to the final design matrix for a total of 13 covariates. We allow an AR order up to a maximum of $P=12$ when fitting the SVARO model, and an $AR(3)$ for the GLM-AR model using both the PMCMC and PVB algorithms. While the choice of an $AR(3)$ for the latter two approaches may seem arbitrary, this is exactly the justification for the use of the SVARO model where such an arbitrary assumption need not be made. We consider data collected on a single subject in what follows. 

Pre-processing steps are applied to the data prior to fitting the Bayesian models. All functional images are aligned to the first image using a six-parameter rigid-body transformation. Then slice-timing correction is performed to set the standard acquisition time as the 12th slice. Images are spatially normalized to a standard EPI image. The global mean, $g$, is computed and each time series is divided by $100/g$ to represent a percentage of $g$. Finally, a high-pass filter with cut-off frequency of $1/128$Hz is apply to the data and design matrix to remove low frequency signals that arise through scanner drift.

\begin{figure}[!ht]
\begin{center}
    \includegraphics[width=\linewidth]{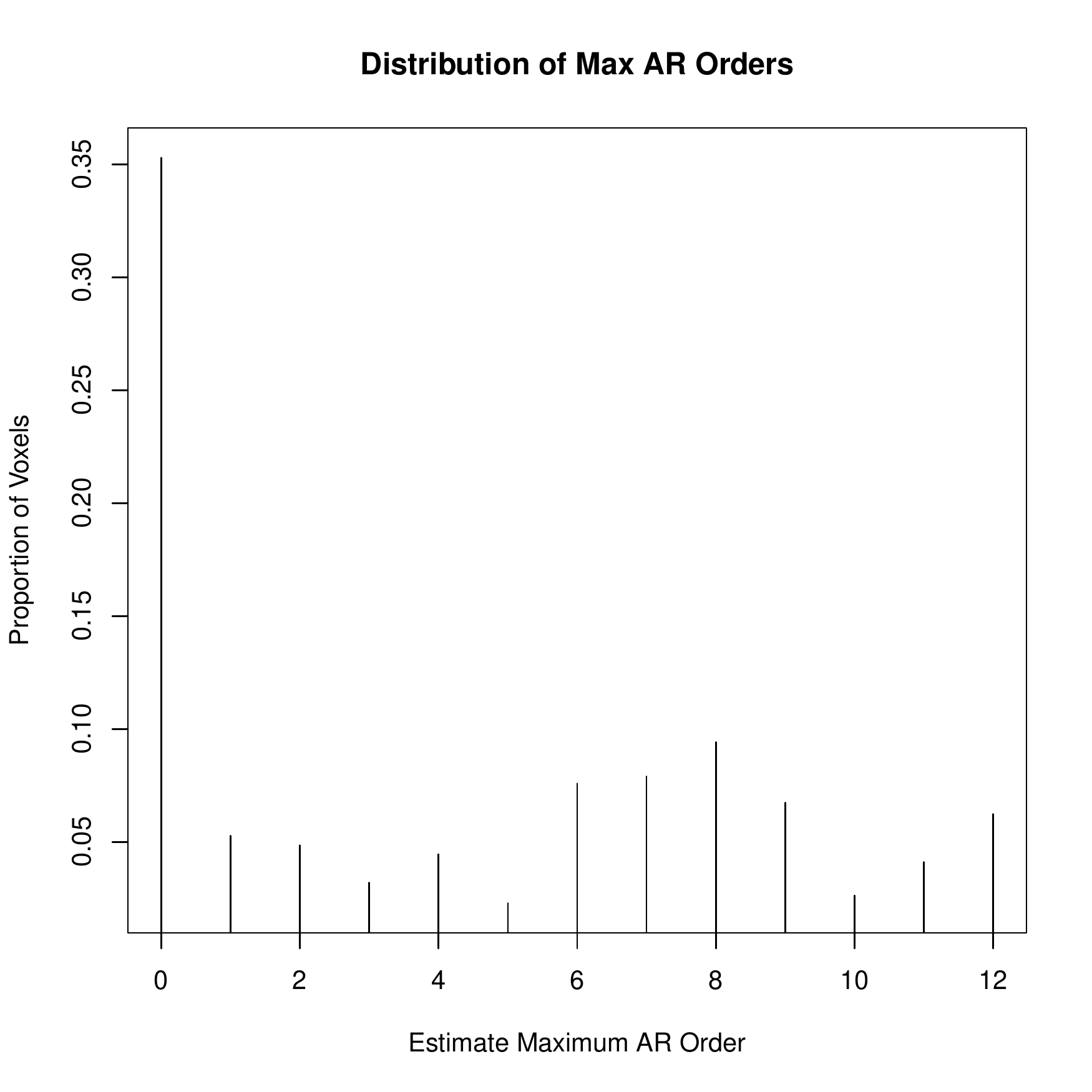}
    \caption{Histogram of the estimated maximum AR order from SVARO across all voxels for the face repetition data set.}
\label{maxARorder}
\end{center}
\end{figure}

Figure \ref{maxARorder} presents the distribution of optimal AR orders estimated from SVARO across voxels. The most frequent order is the zero order, or no autocorrelation in the time series, accounting for approximately 35\% of the voxels.  Interestingly, the next highest is order 8, with 9.4\% of the voxels. Overall, roughly 51\% of the voxels exhibit an AR order greater than 3.  The existence of these higher orders and the variability in the orders is in general agreement with our exploratory analysis of the face repetition data set (see Section \ref{motexamp}) and indicates the necessity of our proposed model. 

\begin{figure}[!ht]
\begin{center}
\includegraphics[width=\linewidth]{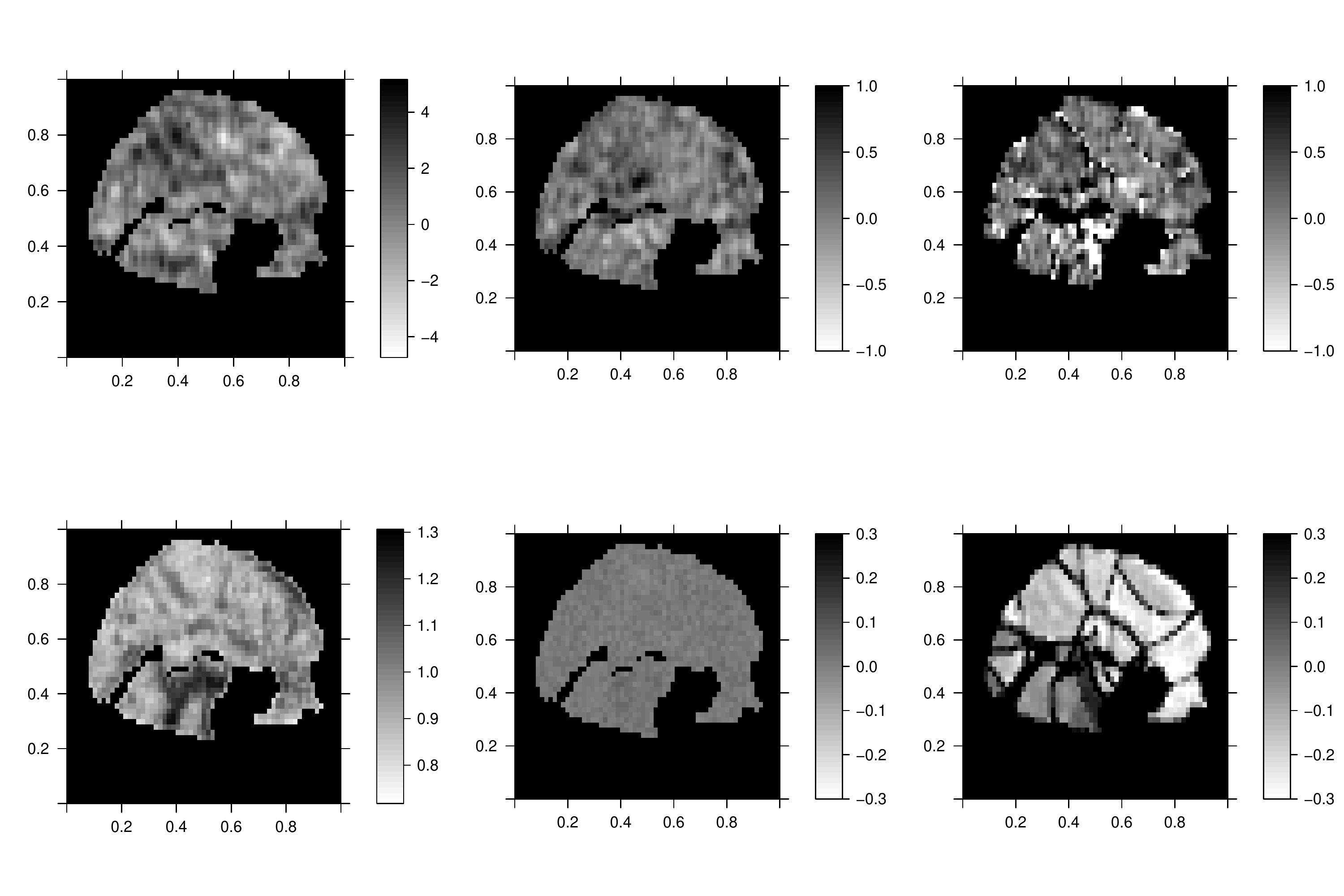}
\caption{Posterior estimates from the middle (27/53) slice of the brain, sagittal view. Top row: posterior mean of fame. Bottom row:  posterior standard deviation of fame. Left column: results from SVARO.  Center column: difference map, SVARO$-$PMCMC. Right column: difference map, SVARO$-$PVB.}
\label{fig_image_real}
\end{center}
\end{figure}

Next, we compare models/algorithms with respect to a particular contrast of interest, namely the effect of ``fame''.  That is, famous faces versus unfamiliar faces.
The estimated posterior mean and standard deviation (SD) maps for the fame contrast, as well as the estimated posterior mean for the $1^{st}$ order autoregressive coefficient, are displayed in Figure \ref{fig_image_real}. While the posterior SD estimated from SVARO and PMCMC are very close, the estimated posterior means are different between the two approaches. Also, the estimated SD obtained from PVB shows apparent discretization. This is due to a graph-partitioning that is incorporated in the algorithm for the sake of computational speed \citep{penny2007bayesian}. It is clear that the boundaries of these graph-partitioned regions have substantially larger estimated SD than the interior locations.  The posterior mean of PVB also seems to exhibit artifacts at the partition boundaries, thought the effect is not as pronounced. The estimated LPML is $-4.66 \times 10^{7}$ under our proposed model and is $-4.82 \times 10^7$ under GLM-AR with MCMC sampling.  According to this model selection criterion,  our proposed model is preferred.

\begin{figure}[!ht]
\begin{center}
\begin{tabular}{ccc}
SVARO & PMCMC & PVB \\
\includegraphics[width=0.3\linewidth]{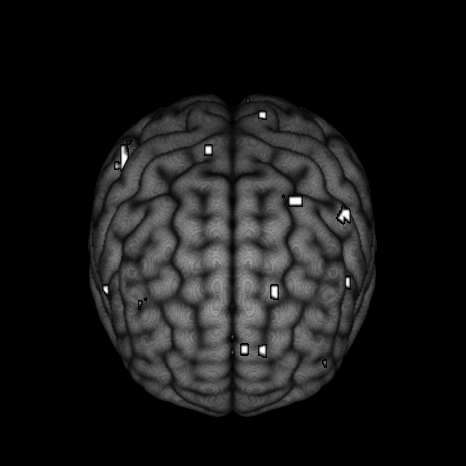}  &
\includegraphics[width=0.3\linewidth]{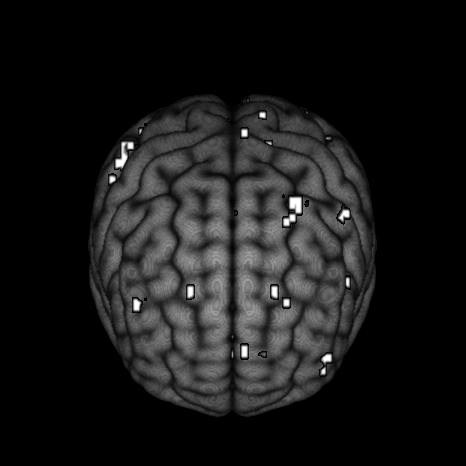} &
\includegraphics[width=0.3\linewidth]{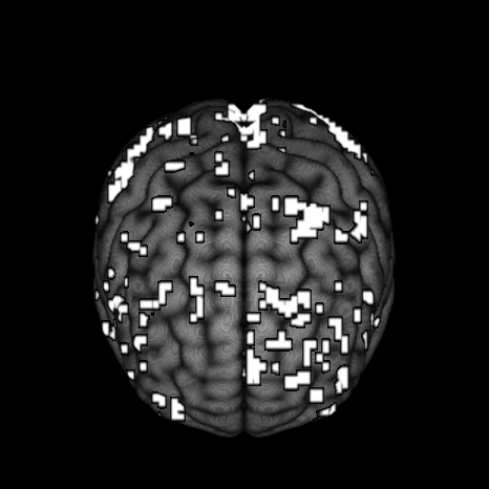}  \\
\includegraphics[width=0.3\linewidth]{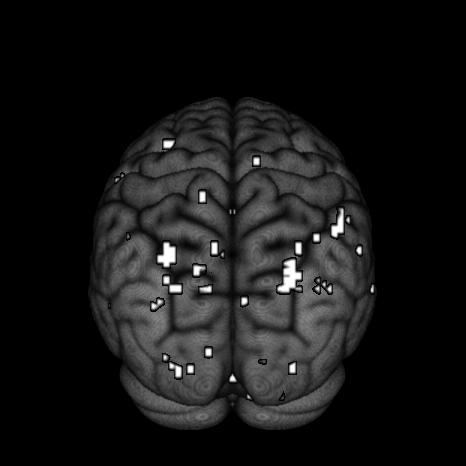} &
\includegraphics[width=0.3\linewidth]{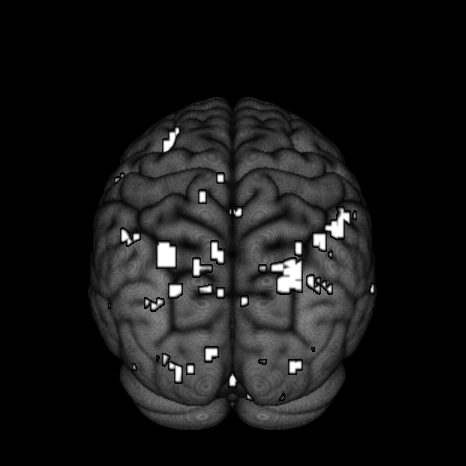} &
\includegraphics[width=0.3\linewidth]{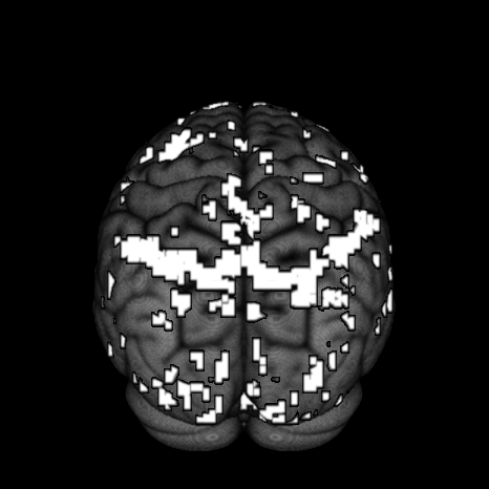} 
\end{tabular}
\caption{Activation maps for fame contrast. Top row: anterior view. Bottom row: posterior views Left column: activation (white colors) from SVARO. Middle column: activation from PMCMC. Right column: activation from PVB.}
\label{fig_fam_3d_real}
\end{center}
\end{figure}

Finally, we look at the effect of fame using thresholded PPMs. The activation threshold is set to $0.2\%$ of the global mean value, and the probability threshold is set to $0.95$. 
Figure \ref{fig_fam_3d_real} shows the activation regions projected onto the pial surface.  We can see that there is a match in terms of a majority of activation regions inferred from SVARO and PMCMC. A closer look reveals that PMCMC tends to make more scattered predictions across the posterior regions of the  brain. The number of activation regions from PVB are far greater than the number obtained from the other two approaches, and are more widely dispersed across the brain. From our simulation study results, we suspect that these scattered activated regions are likely false positives.

In terms of computation time, PVB took about $1$ hour, PMCMC took $1$ day, while SVARO took about $1$ week of computation. Much of our MCMC algorithm is amenable to parallel programming, which is an avenue for further development.

%\begin{figure}[!ht]
%\begin{center}
%\includegraphics[width=\linewidth]{fig_fam_1.png} 
%\includegraphics[width=\linewidth]{fig_fam_2.png} 
%\caption{Activation maps for effect of fame on the three middle slices. From left to right are sagittal, coronal and transverse slice. Top row shows the activation from SVARO (red) and PMCMC (blue), with a joint region indicated by purple dots. The bottom row shows the activation from SVARO (red) and PVB (green), with the joint region denoted by brown dots.}
%\label{fig_fam_real}
%\end{center}
%\end{figure}

\section{Discussion}\label{dis}
In this paper, we have developed a Bayesian hierarchical model, SVARO, that allows the AR order to vary spatially across the brain, with the orders themselves displaying a certain level of spatial clustering based on an Ising model. We compared our proposed model with a self-written MCMC sampler for the standard GLM-AR model and the VB implementation for the same model. The results are interesting, under a low SNR ratio, VB seems to suffer from variance overestimation, leading to a much bigger MSE than the other two methods. It is likely that as temporal noise increases, a more vital role is played by the AR correlation that increases the posterior correlation between different parameters and this makes the mean field assumption underlying the VB approximation less accurate.

We have shown that our model outperforms the GLM-AR model not only in terms of accuracy and sensitivity, but also according to formal model selection based on the LPML criterion. Through an application of our proposed model and through an exploratory analysis, we have shown that the typical constant low-order AR assumption can be violated with real fMRI data. It is very likely that this issue, seen in the face repetition data set, is also present in other fMRI data sets.

There is a computational price to be paid for gaining the flexibility we have proposed in our model. Our model takes a longer time to run than either implementation of GLM-AR. This is mostly due to the estimation of the varying AR orders and the associated greater number of parameters to estimate. However, as previously mentioned, there are elements of our MCMC algorithm that are amenable to parallel programming.  This will be investigated in future work.

While we have based our model specification on a set of independent Ising processes, one for each possible order of the AR process, another approach would be to assume a Potts model for the orders of the AR coefficients. A Potts model, combined with a Dirichlet process prior for parameters has been investigated for selecting covariates of interest in brain imaging \citep{johnson2013bayesian}. Here we can also apply it to the selection of autoregressive orders to yield a more parsimonious, yet still flexible,  model. Investigation of hyper-parameter estimation in the Ising model and the use of alternative spatial models is also of interest, as is increasing the scope of our comparison of methods to include wavelet approaches that focus on long memory errors, or VAR models \citep{harrison2003multivariate}.

%\newpage
\bibliographystyle{rss} 
\bibliography{svaro}

\end{document}